\documentclass[journal=jacsat,manuscript=article]{achemso}

\usepackage[version=3]{mhchem}

\usepackage{achemso}

\usepackage{times}

\usepackage{color, soul}
\usepackage{textcomp}

\usepackage{graphicx}
\graphicspath{ {Figures/} }
\usepackage[table]{xcolor}
\usepackage[labelfont=bf]{caption}
\usepackage{amsmath} 
\usepackage{amssymb} 
\usepackage{bm}

\author{Himanshu Bangar}
 \affiliation{Department of Physics, Indian Institute of Technology Delhi, Hauz Khas, New Delhi 110016, India.}

\author{Pratik Sahu}
 \affiliation{Department of Physics, Indian Institute of Technology Madras, Chennai 600036, India.}
\alsoaffiliation{Center for Atomistic Modelling and Materials Design, Indian Institute of Technology Madras, Chennai 600036, India } 

\author{Akash Kumar}
\affiliation{Department of Physics, University of Gothenburg, Gothenburg 412 96, Sweden.}
\alsoaffiliation{Research Institute of Electrical Communication (RIEC) and Center for Science and Innovation in Spintronics (CSIS), Tohoku University, 2-1-1 Katahira, Aoba-ku, Sendai 980-8577 Japan}
 
\author{Pankhuri Gupta}
\affiliation{Department of Physics, Indian Institute of Technology Delhi, Hauz Khas, New Delhi 110016, India.}

\author{Aman Saxena}
\affiliation{Department of Physics, Indian Institute of Technology Delhi, Hauz Khas, New Delhi 110016, India.}

\author{Sheetal Dewan}
\affiliation{School of Interdisciplinary Research, Indian Institute of Technology Delhi, Hauz Khas, New Delhi-110016, India.}

\author{Samaresh Das}
\affiliation{Center for Applied Research in Electronics, Indian Institute of Technology Delhi, Hauz Khas, New Delhi-110016, India.}

\author{Johan \AA kerman}
\affiliation{Department of Physics, University of Gothenburg, Gothenburg 412 96, Sweden.}
\alsoaffiliation{Research Institute of Electrical Communication (RIEC) and Center for Science and Innovation in Spintronics (CSIS), Tohoku University, 2-1-1 Katahira, Aoba-ku, Sendai 980-8577 Japan}

\author{Birabar Ranjit Kumar Nanda}
\email{nandab@iitm.ac.in}
 \affiliation{Department of Physics, Indian Institute of Technology Madras, Chennai 600036, India.}
 \alsoaffiliation{Center for Atomistic Modelling and Materials Design, Indian Institute of Technology Madras, Chennai 600036, India } 

\author{Pranaba Kishor Muduli}
\email{muduli@physics.iitd.ac.in}
\affiliation{Department of Physics, Indian Institute of Technology Delhi, Hauz Khas, New Delhi 110016, India.}

\title{Giant Damping-like Spin-Torque Conductivity in a GeTe/Py van der Waals Heterostructure}

\begin{document}

\newpage

\begin{abstract}
Recent observations of large unconventional spin-orbit torques in van der Waals (vdW) materials are driving intense interest for energy-efficient spintronic applications. A key limitation of ferromagnet (FM)/vdW heterostructures is their lower value of damping-like torque conductivity ($\sigma{\rm_{DL}^{y}}$) compared to the conventional heavy metal-based systems, limiting their prospects for commercial spintronic devices. Here, we report both a giant $\sigma{\rm_{DL}^{y}}$ of $-(1.25 \pm 0.11)\times 10^{5}~\hbar/ 2e~\Omega^{-1}$m$^{-1}$ and an unconventional spin-orbit torque in a heterostructure comprising an FM (Ni$_{80}$Fe$_{20}$) and the vdW material GeTe. The value of $\sigma{\rm_{DL}^{y}}$ represents the highest reported torque conductivity for any FM/vdW interface and is comparable to benchmark heavy metal heterostructures. First-principles calculations reveal that this substantial torque originates from the cooperative interplay of the spin Hall effect, orbital Hall effect, and orbital Rashba effect, assisted by interfacial charge transfer. These findings demonstrate the potential of carefully engineered vdW heterostructures to achieve highly efficient electrical manipulation of magnetization at room temperature, paving the way for next-generation low-power spintronic devices.

\end{abstract}


\section{INTRODUCTION}

Spintronics is an emerging field that harnesses not only the charge of electrons but also their intrinsic spin degree of freedom to develop efficient, high-speed, and scalable memory and logic technologies.~\cite{hirohata2020review} A key aspect of spintronic devices is the effective generation and manipulation of spin currents.~\cite{sinova2015spin} These spin currents can exert a spin-orbit torque (SOT) on an adjacent ferromagnetic (FM) layer, enabling magnetization switching and auto-oscillations, which are essential for data storage and processing.~\cite{myers1999current,miron2011perpendicular,kumar2025spin} The SOT offers substantial advantages in energy efficiency and endurance for advanced non-volatile memory and logic technologies.~\cite{dieny2020opportunities} Spin currents needed for SOT devices are typically generated through mechanisms such as the spin Hall effect (SHE)~\cite{hirsch1999spin,dyakonov1971current,sinova2015spin,bangar2023large} or the Rashba-Edelstein effect (REE)~\cite{edelstein1990spin,bychkov1984properties,shao2016strong,bangar2022large}. 


Recently, an orbital counterpart to both the SHE and REE effects, known as the orbital Hall effect (OHE) and the orbital Rashba effect (ORE), has been discovered.~\cite{jo2018gigantic,park2011orbital,cysne2025orbitronics,chen2018giant,krishnia2024quantifying,el2023observation} 
These orbital effects offer an additional mechanism for spin-charge conversion and can significantly enhance SOT efficiency.~\cite{yang2024harnessing,el2023observation}
Until now, most of the studies on OHE and ORE have focused on light elements with low spin-orbit coupling (SOC), such as Ti, Cu, and Al.~\cite{go2018intrinsic,kontani2009giant,rothschild2022generation,choi2023observation,lyalin2023magneto,lee2021efficient} In these materials, OHE is dominant over SHE.~\cite{choi2023observation,an2023electrical} On the other hand, in heavy metals (HMs) with strong SOC, both the OHE and the SHE are significant. Recent studies show that the OHE is even more pronounced in HMs, serving as a fundamental generator of angular momentum. It is now believed that SHE in these HMs partially results from the conversion of the orbital current into a spin current through SOC.~\cite{sala2022giant} Such combinations of orbital and spin current can enhance the strength of SOT and improve the performance of spintronic devices.~\cite{yang2024harnessing} However, these traditional HMs, as well as the light metals, face a fundamental limitation. The inherent in-plane spin polarization of the spin current generated in these materials does not provide the necessary deterministic switching of the FM layer when used with the perpendicular magnetic anisotropy required for high-density magnetic recording.~\cite{fukami2016spin,shao2021roadmap,manchon2019current}

Two-dimensional van der Waals (vdW) materials and interfaces represent a class of materials that have garnered substantial interest for their application in SOT-based devices~\cite{liu2020two,kurebayashi2022magnetism}. For example, the low-crystalline symmetry-based vdW systems have demonstrated the capability to generate out-of-plane damping-like torques,~\cite{macneill2017control,bainsla2024large,macneill2017thickness,kao2022deterministic,shi2019all,shin2022spin} crucial for efficiently switching perpendicularly magnetized systems, particularly in high-density magnetic recording applications.~\cite{kao2022deterministic,shi2019all,shin2022spin} Several recent studies have systematically explored a variety of transition metal dichalcogenides-based heterostructures, such as MoS$_2$, MoTe$_2$, and WS$_2$ in combination with FMs, uncovering robust SOT as well as highly tunable spin-charge interconversion mechanisms~\cite{shao2016strong, zhang2016research, macneill2017control, macneill2017thickness, guimaraes2018spin, stiehl2019layer}. 
Despite these advances, the damping-like conductivity in vdW/FM-based heterostructures remains lower compared to conventional HM-based systems. Recent theoretical works showing large orbital currents in vdW materials\cite{sahu2024emergence,cysne2025orbitronics}  present an opportunity to enhance damping-like conductivity, a topic that has not been extensively explored. 

GeTe is a notable vdW material that has not yet been explored in the context of OHE. GeTe exhibits a non-centrosymmetric rhombohedral crystal structure with space group $R3m$ at room temperature [Fig.~\ref{fig:1}(a)]~\cite{shelimova1993crystal}. Its structure is distorted approximately 1.65\textdegree along the [111] direction, resulting in three longer (3.15 $\rm \AA$) and three shorter (2.83 $\rm \AA$) bonds, characteristic of Peierls distortion [Fig.~\ref{fig:1}(a)]~\cite{fons2010phase,chatterji2015anomalous}. This distortion breaks inversion symmetry and, along with SOC, causes Rashba splitting of bulk bands, as illustrated in Figure~\ref{fig:1}(b)~\cite{di2012electric, liebmann2016giant}.  The bulk Rashba parameter, denoted as $\alpha _{\rm R}$, measuring  4.2 eV\AA $-$one of the highest values known for any material~\cite{krempasky2016disentangling}. Theoretical calculations also indicate the presence of the spin Rashba effect (SRE) in GeTe due to the large Rashba splitting of bulk bands.~\cite{zhang2020tuning, wang2020spin}. Furthermore, based on computational studies of similar two-dimensional compounds~\cite{sahu2024emergence}, the ORE is also strongly anticipated in GeTe [schematically represented in Fig. \ref{fig:1}(c)]. The combined influence of SRE and ORE suggests that GeTe-based heterostructures should generate substantial SOT.
Furthermore, low crystalline symmetry of GeTe is predicted to generate out-of-plane polarized spin current ($\sigma _{\rm z}$)~\cite{liu2021symmetry} needed for high-density magnetic recording. In fact, a recent work by Jeon \textit{et al.} reported the presence of an unconventional field-like torque in the epitaxial GeTe/Ni$_{80}$Fe$_{20}$ bilayer~\cite{jeon2021field}. However, a large damping-like torque, essential for efficient SOT switching in GeTe-based systems, is yet to be reported. 

In this article, we investigated SOT in GeTe/Ni$_{80}$Fe$_{20}$ (hereby Ni$_{80}$Fe$_{20}$ is referred to as Py) bilayer heterostructures. We grew polycrystalline thin films of GeTe on CMOS-compatible Si(111) substrates using the pulsed laser deposition technique. Subsequently, we studied SOT in GeTe/Py bilayer-based devices. 
We demonstrate both unconventional SOT and a remarkably giant damping-like torque conductivity $\sigma_{DL}^{y}$ 
of $-(1.2 \pm 0.1)\times 10^{5}~\hbar/ 2e~\Omega^{-1}$m$^{-1}$. 
This value represents the highest reported torque conductivity for any FM/vdW interface to date and is comparable to what has been achieved in benchmark HM heterostructures. Our first-principles calculations indicate that this significant damping-like torque arises from the cooperative interplay of the SHE, OHE, and ORE within GeTe, with additional contributions from interfacial charge transfer. These findings collectively highlight the tremendous potential of carefully engineered vdW heterostructures for enabling highly efficient electrical manipulation of magnetization at room temperature, thereby paving the way for next-generation, low-power spintronic devices.

\section{RESULTS AND DISCUSSION}

\subsection{Growth of $\alpha$-GeTe thin films}

\begin{figure*}[t!]
\includegraphics[width=\textwidth]{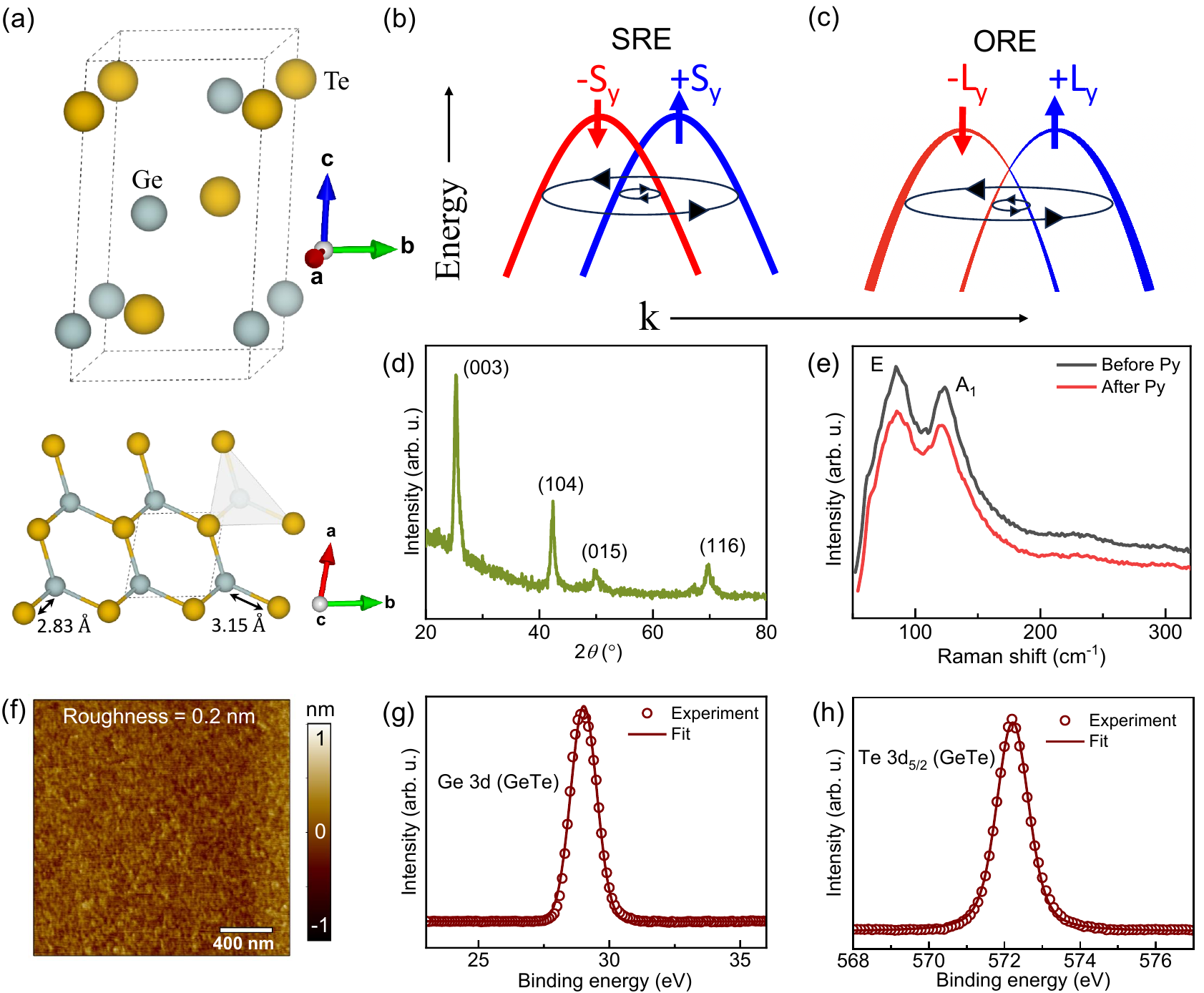}
\caption{ \textbf{Crystal structure and characterization of GeTe thin films.} (a) Crystal structure of $\alpha$-GeTe (rhombohedral structure, $R3m$). The lower panel illustrates the GeTe structure viewed along the $c-$axis, where the Ge and Te atoms are bonded by three shorter (2.83 $\textup{\AA}$) and three longer (3.15 $\textup{\AA}$) bonds due to Peierls distortion. 
Schematic illustration of the splitting of bulk bands in GeTe due to (b) spin Rashba effect (SRE) and (c) orbital Rashba effect (ORE). (d) Glancing angle x-ray diffraction pattern of a 23 nm thick GeTe thin film, confirming the phase purity and match with the $\alpha$-phase of GeTe. (e) The room-temperature Raman spectra before (black) and after (red) deposition of Py on top of GeTe, showing the characteristic $E$ and A$_1$ modes at 88.3 and 125.6 cm$^{-1}$. (f) Atomic force microscopy image of GeTe thin film, demonstrating a smooth surface morphology with an root mean square roughness of $\approx$~0.2 nm. (g)-(h) X-ray photoelectron spectroscopy spectra showing the core level peaks for Ge 3d and Te 3d$_{5/2}$ peaks, respectively, for GeTe thin film capped with Al. The solid lines represent the fitted data. 
}
\label{fig:1}
\end{figure*}

We optimized the growth of polycrystalline $\alpha$-GeTe on a Si(111) substrate using pulsed laser deposition. Figure~\ref{fig:1}(d) shows the grazing incidence x-ray diffraction pattern of a 23~nm-thick GeTe thin film [capped with Al (3 nm)] grown at 250~$^\circ$C. The diffraction peaks at angles of 25.1\textdegree,~42.2\textdegree,~49.7\textdegree~and 69.5\textdegree~corresponds to the lattice planes (003), (024), (205) and (226)
of rhombohedral GeTe, respectively. These Bragg reflection peaks were successfully indexed to the  rhombohedral crystal structure of GeTe with the space group $R3m$ (JCPDS no. 47-1079)~\cite{kalra2015role}. This x-ray diffraction spectrum confirms the growth of a polycrystalline $\alpha$-GeTe thin film without any impurity phases.

In addition to x-ray diffraction, we conducted Raman spectroscopy on the 23 nm thick GeTe thin film [black curve in Fig.~\ref{fig:1}(e)], revealing the characteristic modes E and A$_1$ of $\alpha$-GeTe at 83.8 cm$^{-1}$ and 122.6 cm$^{-1}$~\cite{wang2016ordered}. To evaluate the surface morphology, atomic force microscopy measurements were performed. Figure~\ref{fig:1}(f) shows the atomic force microscopy image of the GeTe thin film, revealing a smooth surface with a root-mean-square roughness of $\approx$~0.2 nm.

Further, we conducted x-ray photoelectron spectroscopy (XPS) measurements to examine the chemical states of Ge and Te. GeTe is found to easily oxidize under atmospheric conditions as confirmed by depth-profiled measurements (refer to Section S1 in the supplementary information). Hence, we capped GeTe films with a 2 nm-AlO$_{\rm x}$. Figures~\ref{fig:1}(g) and (h) show the XPS spectra measured at room temperature for the Ge 3d and Te 3d$_{5/2}$ peaks, respectively, with the capping layer. The solid lines in the figures represent the peak fits. The Ge 3d peak at the binding energy of 29.01 $\pm$ 0.01 and the Te 3d$_{5/2}$ at 572.22 $\pm$ 0.01 matches well with the literature.~\cite{yashina2001xps} The XPS data confirmed the core levels of Ge and Te without any signs of oxidation in our samples.

To form a heterostructure with a FM for investigating SOT, we transfer the GeTe/AlO$_{\rm x}$ samples to a separate sputtering system. We deposited the Py layer by removing the entire AlO$_{\rm x}$ capping layer on GeTe and around 8 nm of the GeTe layer using substrate bias at 2$\times 10^{-3}$~Torr Ar pressure. This process creates a sharp interface between the GeTe and Py. The parameters used for substrate bias are same as in one of our earlier work.~\cite{agarwal2023strong}. A 2 nm Al layer was deposited as the capping layer on top of GeTe/Py. 
To assess the impact of Py deposition on the quality of GeTe film, we examined the Raman spectra, as depicted in Fig.~\ref{fig:1}(e). The Raman data clearly indicates that the GeTe layer remains intact, and there is no damage resulting from the Py deposition. Subsequently, standard optical lithography and Ar-ion milling techniques were employed to fabricate GeTe/Py microstrip devices (see methods for more details).

\subsection{Spin-orbit torques in GeTe/Py}

\begin{figure*}[t]
\includegraphics[width=\textwidth]{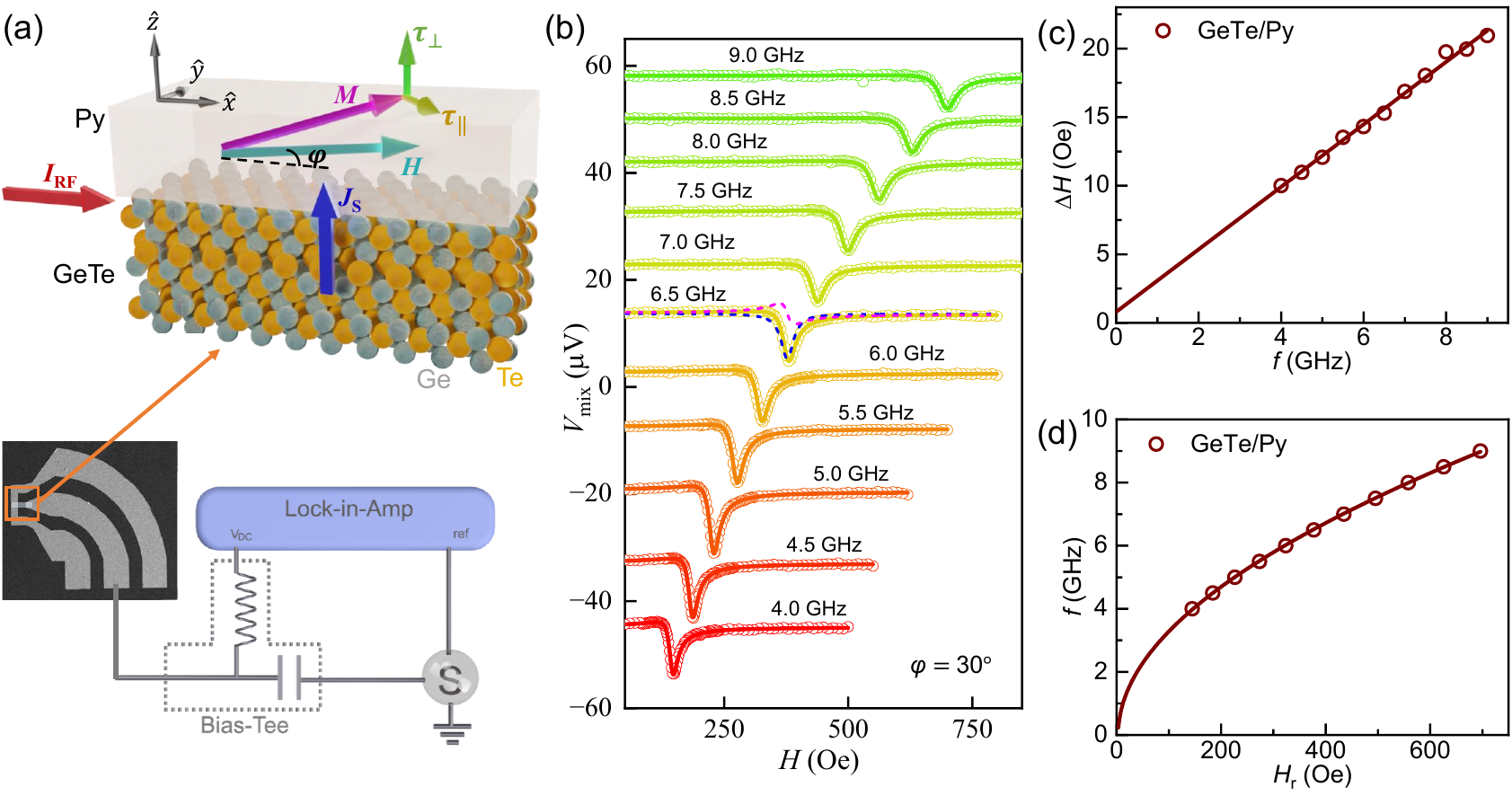}
\caption{\textbf{Spin-torque ferromagnetic resonance measurements in GeTe/Py heterostructures.} (a) The schematic of the device consisting of a GeTe/Py bilayer, the setup for STFMR measurements, and a scanning electron microscopy image of the microstrip device with a co-planar waveguide. The top panel shows the schematic of the generation of both in-plane ($\tau _{\parallel}$) and out-of-plane ($\tau _{\bot}$) torque when the device is subject to electric current ($I_{\rm RF}$) flowing along the $x$-direction. (b) The frequency-dependent STFMR spectra (open symbols) and their fits (solid line) for the GeTe(15 nm)/Py(8 nm) sample.  An example of decomposition of fitting into the symmetric (blue dotted curve) (V$_{\textrm{S}}$) and anti-symmetric (V$_{\textrm{A}}$) component (magenta dotted curve)  is shown for 6.5 GHz. (c) Linewidth ($\Delta H$) vs. frequency ($f$) (open circle) and their corresponding fit (solid lines). (d) $f$ vs. $H_r$ curve (open circle) and their corresponding fit (solid lines) with Kittel's equation.}
\label{fig:2}
\end{figure*}

We investigate the SOTs in GeTe(15 nm)/Py(8 nm) by performing spin-torque ferromagnetic resonance (STFMR) measurements. The device structure and experimental setup are illustrated in detail in Fig.~\ref{fig:2}(a). The microstrip device, which consists of a GeTe/Py bilayer, is shown in the scanning electron microscopy image. This device is configured with a co-planar waveguide to deliver the microwave current for STFMR measurements, wherein we apply an RF current ($I_{\rm RF}$) along the $x-$direction and a magnetic field ($H$) at an angle $\varphi$, relative to $I_{\rm RF}$. 
The STFMR spectra ($V_{\rm mix}$) at different frequencies are shown in Fig.~\ref{fig:2}(b), which is dominated by a large symmetric component in contrast to  Jeon \textit{et al.} ~\cite{jeon2021field}. The lineshape of STFMR can be effectively fitted with a combination of symmetric and anti-symmetric Lorentzian function:~\cite{liu2011spin,You2021}

\begin{equation}\label{Vmix}
 {V_{\rm mix}} = {V_{\rm S}} \frac{\Delta H^2}{\Delta H^2+(H-H_{\rm r})^2}\\+{V_{\rm A}} \frac{\Delta H(H-H_{\rm r})}{\Delta H^2+(H-H_{\rm r})^2}
\end{equation}

where, $\Delta H$ and $H_{\rm r}$ are the linewidth and resonance field, respectively. The symmetric ($V_{\rm S}$) and antisymmetric ($V_{\rm A}$) components are proportional to the in-plane ($\tau _{\parallel}$) and out-of-plane ($\tau _{\bot}$) torques, respectively. The $\Delta H$ vs. frequency ($f$) dependence is shown in Fig.~\ref{fig:2}(c) and fitted using the following equation: $\Delta H=\frac{2\pi\alpha_{\rm eff} f}{\gamma}+\Delta H_{\rm 0}$. Here, $\alpha _{\rm eff}$ is the effective Gilbert damping parameter, $\gamma = 1.85 \times 10^2 $ GHz/T is the gyromagnetic ratio, and $\Delta H_{\rm 0}$ is the inhomogeneous line broadening. The first term on the right-hand side is the viscous damping of magnetization motion, while the second term is due to magnetic inhomogeneity and sample imperfections of the FM layer~\cite{farle1998ferromagnetic}. The slope of the linear fit is proportional to $\alpha _{\rm eff}$, which came out to be 0.0067 $\pm$ 0.0002. Figure~\ref{fig:2}(d) shows the $f$ dependent $H_{\rm r}$ (open circles) with its corresponding fit (solid line) using Kittel's equation~\cite{kittel1948theory}. From the fit, the effective magnetization ($M_{\rm eff}$) is found to be  1.26 $\pm$ 0.01 T. The $M_{\rm eff}$ in our Py sample is close to the previously reported values showing good quality of Py.\cite{de2007ferromagnetic}


\begin{figure*}[t]
\includegraphics[width=0.85\textwidth]{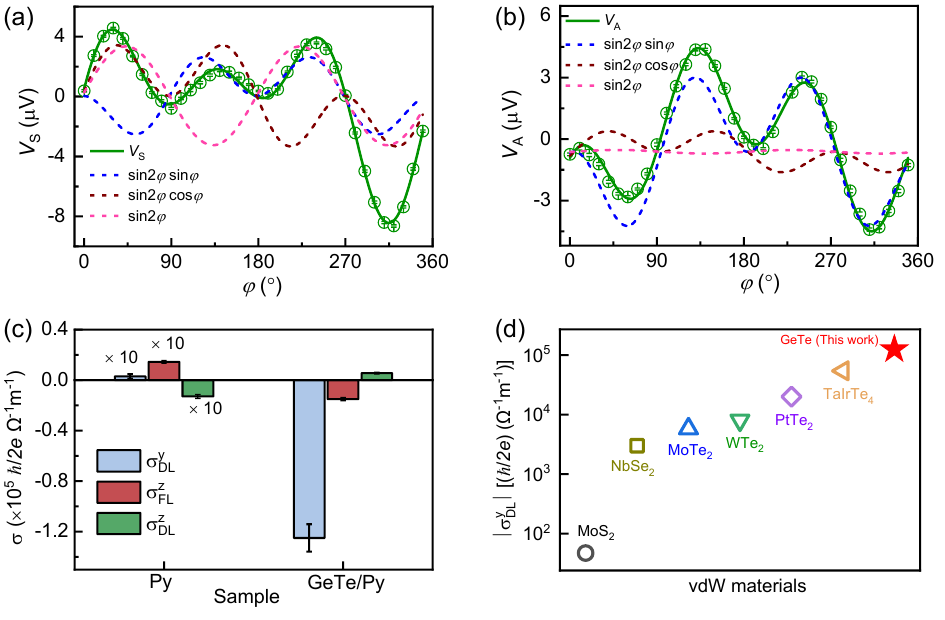}
\caption{\textbf{Angle-dependent STFMR measurements and comparison of spin Hall conductivity in vdW Heterostructures.} The angular dependence of (a) Symmetrical component, V$_{\textrm{S}}$ and (b) Asymmetrical component, V$_{\textrm{A}}$ of STFMR for the GeTe/Py sample. Experimental data (open circles) are fitted (solid lines) using $\text{Eq.}~\ref{VS}$ in panel (a) and $\text{Eq.}~\ref{VA}$ in panel (b). The dashed lines indicate the angular dependence of individual spin-torque contributions. (c)  Extracted spin-torque conductivities ($\sigma_{\rm DL}^{y}$, $\sigma_{\rm FL}^{z}$, and $\sigma_{\rm DL}^{z}$) for the GeTe/Py together with a reference Py device without the GeTe layer. The spin-torque conductivities for the Py device are scaled by $\times 10$ to enhance their visualization relative to GeTe/Py. (d) Comparative analysis of the dominant damping-like spin-torque conductivity $\sigma_{\rm DL}^{y}$ for $\alpha$-GeTe (this work) against other vdW materials: MoS$_2$~\cite{safeer2019room}, MoTe$_2$~\cite{stiehl2019layer}, WTe$_2$~\cite{macneill2017control}, NbSe$_2$~\cite{guimaraes2018spin}, PtTe$_2$~\cite{xu2020high}, and TaIrTe$_4$~\cite{liu2023field}.}
\label{fig:3}
\end{figure*}

The SOTs corresponding to different components of spin polarization can be obtained from the analysis of in-plane magnetic field angle ($\varphi$) dependent STFMR measurements. Figure~\ref{fig:3}(a) and (b) presents the angle-resolved V$_{\textrm{S}}$ and V$_{\textrm{A}}$ components, respectively. For conventional systems involving HM such as Pt and Ta,~\cite{macneill2017control} where the spin polarization is oriented along the $y-$axis ($p _{y}$), both the damping-like torque [$\tau ^{y}_{ \rm DL} \propto \hat{ m} \times (\hat{ m} \times \hat{y})$] and the field-like torque [$\tau ^{y}_{ \rm FL} \propto \hat{ m} \times \hat{y}$] exhibit a cos$(\varphi)$ dependence. Additionally, the anisotropic magnetoresiatnce (AMR) in Py, proportional to $\sin(2\varphi)$, contributes to an overall angular dependence of $\sin(2\varphi)\cos(\varphi)$ for both $V_{\rm S}$ and $V_{\rm A}$. However, if the spin polarization differs from $p_{y}$, additional unconventional torques may come into play~\cite{yu2014switching,macneill2017control,bangar2023optimization}. Torques arising from $p _{x}$ [$\tau ^{x}_{\rm DL} \propto \hat{ m} \times (\hat{ m} \times \hat{x})$, $\tau ^{x}_{ \rm FL} \propto \hat{ m} \times \hat{x}$] would generate a $\sin(\varphi)$ dependence, while torques due to $p _{z}$ [$\tau ^{z}_{ \rm DL} \propto \hat{ m} \times (\hat{ m} \times \hat{z})$, $\tau ^{z}_{ \rm FL} \propto \hat{ m} \times \hat{z}$] would be independent of $\varphi$. Therefore, a generalized expression for the angular dependence of $V_{\rm S}$ and $V_{\rm A}$, encompassing all torque components can be formulated as~\cite{zhou2020magnetic,nan2020controlling}:

\begin{equation}\label{VS}
{V_{\rm S}} \propto {\rm sin(2\varphi)} \Bigl( S_{\rm DL}^{x}{ \rm sin(\varphi)} + S_{\rm DL}^{y}{\rm cos(\varphi)}+ S_{\rm FL}^{z} \Bigl)
\end{equation}
\begin{equation}\label{VA}
{V_{\rm A}} \propto {\rm sin(2\varphi)} \Bigl( A_{\rm FL}^{x}{ \rm sin(\varphi)} + A_{\rm FL}^{y}{\rm cos(\varphi)}+ A_{\rm DL}^{z} \Bigl)
\end{equation}

The quantities $S_{\rm DL}^{x}$, $S_{\rm DL}^{y}$, and $A_{\rm DL}^{z}$ are proportional to the damping-like torque generated by distinct components of the spin-torque conductivity tensor, namely $\sigma_{zx}^{x}$, $\sigma_{zx}^{y}$, and $\sigma_{zx}^{z}$, respectively. Similarly, $A_{\rm FL}^{x}$, $A_{\rm FL}^{y}$, and $S_{\rm FL}^{z}$ are proportional to the corresponding field-like component of the spin-torque conductivity tensor. The superscript indicates the direction of spin polarization in the presence of a charge current along the $x-$direction.

\subsection{Quantification of spin-orbit torques in GeTe/Py}
The angle-dependent V$_{\textrm{S}}$ and V$_{\textrm{A}}$ components with the fits are shown in Fig.~\ref{fig:3}(a) and (b), respectively. The fitting parameters $S_{\rm DL(FL)}^{y(z)}$ and $A_{\rm FL(DL)}^{y(z)}$ are correlated with torques $\tau_{\rm DL(FL)}^{y(z)}$ and $\tau_{\rm FL(DL)}^{y(z)}$ through relationships:~\cite{liu2019current, mellnik2014spin, macneill2017control, stiehl2019layer}

\begin{equation}\label{TS}
\tau_{\rm DL(FL)}^{y(z)} = \frac{2 S_{\rm DL(FL)}^{y(z)} \alpha\gamma (2H_{\rm r}+\mu_{\rm 0}M_{\rm eff})}{\Delta R I_{\rm RF}}
\end{equation}
\begin{equation}\label{TA}
\tau_{\rm FL(DL)}^{y(z)} = \frac{2 A_{\rm FL(DL)}^{y(z)} \alpha\gamma (2H_{\rm r}+\mu_{\rm 0}M_{\rm eff})}{\Delta R I_{\rm RF}\sqrt{1+\frac{\mu_{\rm 0} M_{\mathrm{eff}}}{H_{\rm r}}}}
\end{equation}

Here, $\Delta R$ represents the AMR of Py determined from the fitting of resistance as a function of angle $\varphi$. $I_{\rm RF}$ is the microwave current through the device, which can be calculated using the transmitted power, $P_{\mathrm{trans}}$, and the device's impedance, $Z$. This relationship is expressed as $I_\mathrm{RF} = \sqrt{\frac{2 P_{\mathrm{trans}}}{Z}}$. The value of $P_{\mathrm{trans}}$ is determined from the measurement of the S$_{11}$ parameter using a vector network analyzer, which was performed on the same device. Through the parallel resistor model, we found that $\approx22$ \% of $I_\mathrm{RF}$ flows through the GeTe layer (see supplementary section S2).

Finally, we quantify the strength of the individual components of the torques. We define the torque conductivity as nominally independent of geometric factors. It is defined as the amount of angular momentum the magnet absorbs per second, per unit interface area, and per unit electric field. For each torque value, $\tau_{\rm DL, FL}^{y, z}$, the torque conductivity can be defined as:~\cite{macneill2017control, stiehl2019layer} 

\begin{equation}\label{SHC}
\sigma_{\rm DL, FL}^{y, z} = \frac{M_{\rm S}lwt_{\rm Py}}{\gamma}\frac{\tau_{\rm DL, FL}^{y, z}}{(lw)E} = \frac{M_{\rm S}lt_{\rm Py}}{\gamma}\frac{\tau_{\rm DL, FL}^{y, z}}{I_{\rm RF}Z}
\end{equation}

Here $M_{\rm S}$ is the saturation magnetization which is approximated as $M_{\rm eff}$, $E$ is the electric field, $l$ and $w$ are the length and width of the microstrip, $t_{\rm Py}$ is the thickness of the Py layer, and $Z$ is the device impedance. Using Eq.~\ref{SHC}, we found spin torque conductivity~(in units of $10^{5}~\hbar/2e~ \Omega^{-1} \rm{m}^{-1}$) as: $\sigma_{\rm DL}^{y}$ = (-1.25 $\pm$ 0.11), $\sigma_{\rm FL}^{z}$ = (-0.15 $\pm$ 0.01) and $\sigma_{\rm DL}^{z}$ = (0.06 $\pm$ 0.01),  for GeTe/Py sample. Hence, we obtain a large $\sigma_{\rm DL}^{y}$ in direct contradiction to an earlier work by Jeon \textit{et al. } on epitaxial GeTe/Py~\cite{jeon2021field}. We also obtain a significant field-like torque conductivity, $\sigma_{\rm FL}^{z}$, which agrees with the work by Jeon \textit{et al. }~\cite{jeon2021field}

The fitting results reveal the presence of various torque components. The potential origins of these torques include the SHE, OHE, and bulk RE within GeTe, as well as the interfacial Rashba or ORE occurring at the interface between GeTe and Py. To investigate the contributions from GeTe, we compare the $\sigma_{\rm DL}^{y}$ of the GeTe/Py sample with the Py control sample, as depicted in Fig.~\ref{fig:3}(c). 
The $\sigma_{\rm DL}^{y}$ for the Py sample is negligibly smaller than that for the GeTe/Py sample, indicating that both conventional and unconventional SOTs originate from the spin current generated due to the GeTe layer. It is to be noted that $\sigma_{\rm DL}^{y}$ for the Py device are scaled by $\times 10$ to enhance their visualization relative to GeTe/Py in Fig.~\ref{fig:3}(c).

To benchmark our results against widely studied vdW materials, we have plotted the values of $\sigma_{\rm DL}^{y}$ in Fig.~\ref{fig:3}(d). The $\sigma_{\rm DL}^{y}$ values are significantly higher than those of other vdW materials such as MoS$_2$ ($4.7 \times 10^{1}$)~\cite{safeer2019room}, MoTe$_2$ ($5.8 \times 10^{3}$)~\cite{stiehl2019layer}, WTe$_2$ ($8 \times 10^{3}$)~\cite{macneill2017control}, NbSe$_2$ ($3 \times 10^{3}$)~\cite{guimaraes2018spin}, PtTe$_2$ ($0.2-1.6 \times 10^{5}$)~\cite{xu2020high}, and TaIrTe$_4$ ($5.44 \times 10^{4}$)~\cite{liu2023field}, numbers in bracket are in units of $\hbar/ 2e~\Omega^{-1}$m$^{-1}$. Therefore, GeTe heterostructures significantly outperform other vdW material-based heterostructures  with a giant value of $\sigma_{\rm DL}^{y}$. Additionally, the estimated $\sigma_{\rm DL}^{y}$ is in the same order as that of HM such as Pt ($3.4 \times 10^{5}$),~\cite{liu2011spin} $\beta$-Ta ($0.8 \times 10^{5}$),~\cite{liu2012spin} and $\beta$-W ($1.5 \times 10^{5}$)~\cite{sui2017giant} in units of $\hbar/ 2e~\Omega^{-1}$m$^{-1}$.


\section{Theoretical investigation} 
In this section, we present a theoretical perspective on the origin of a large damping-like torque in GeTe/Py heterostructure. The damping-like torque emerges due to (i) transverse conductivities such as OHE~\cite{choi2023observation,lee2021efficient,jo2018gigantic}, SHE~\cite{manchon2019current,sinova2015spin,liu2012spin}, and (ii) ORE~\cite{chen2018giant,krishnia2024quantifying,el2023observation}. While the spin Hall conductivity (SHC) directly contributes to the damping-like torque, the orbital Hall conductivity (OHC) does it via SOC~\cite{yang2024harnessing}. Each of them are sensitive to the Fermi energy, and therefore the charge transfer which often happens in heterostructures due to electronic reconstruction also play an important role. For this purpose, we carried out DFT calculation for a model system with 4 unit cells thick GeTe with a 2 unit cells thick Py overlayer. The details are in section S3 of the Supplementary Material.  We find that there is a net electron transfer of 0.25$e$ per unit cell from GeTe side to Py side. This charge transfer is responsible for a shift in the Fermi energy and hence will have a substantial effect on the SHC and OHC as we will see next.


\begin{figure}[!htb]
    \centering
    \includegraphics[width=\linewidth]{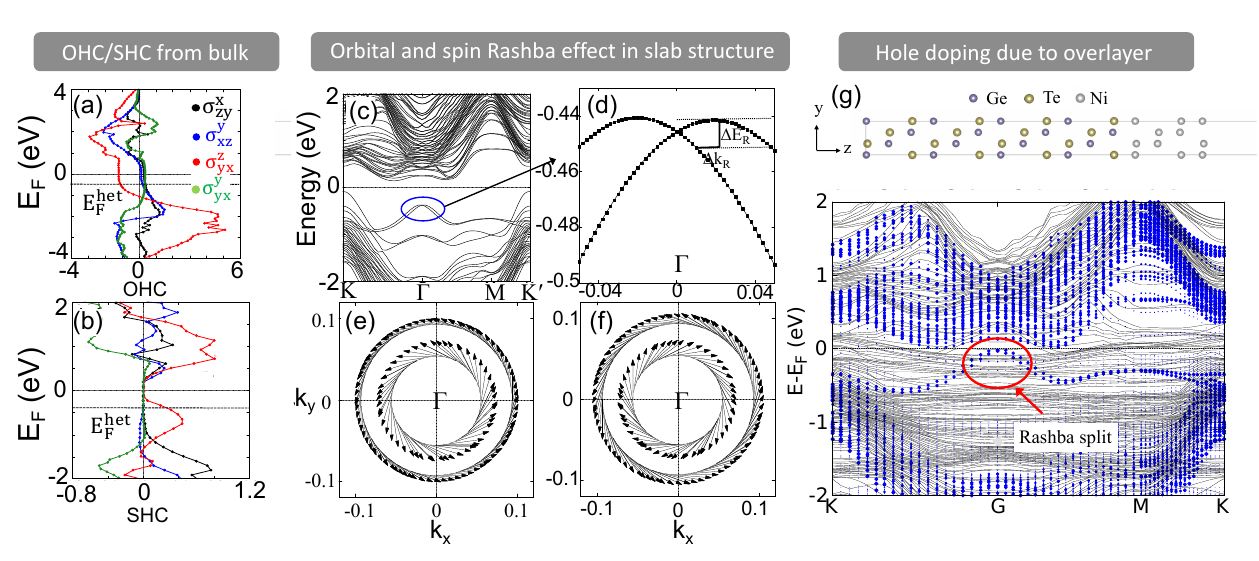}
    \caption{\textbf{Electronic structure, spin and orbital Hall conductivity calculations.} 
    All non zero components of the (a) spin and (b) orbital Hall conductivity tensors are plotted as a function of the Fermi energy. The conductivities are in the units of $10^5 (\hbar/2e) \Omega^{-1} m^{-1}$. The doped Fermi level induced by the Py overlayer is shown by the line $E_F^{het}$. The conductivities at $E_F^\prime$ listed in Table \ref{tab1}.
    (c) Rashba split  near $\Gamma$ point in the band structure for the 4-layer GeTe, (d) energy and momentum shifts for the Rashba splitted bands, (e) orbital texture and (f) spin texture near the $\Gamma$ point. (g) \textit{top} : Structure of the 4-layer GeTe with 4-layers of Ni. Note that here GeTe atoms for the top layer are replaced with the Ni atoms for simplicity; \textit{bottom} : Band structure for the corresponding structure showing the shift of $\Gamma$ point close to Fermi.}
    \label{fig:theory_combined}
\end{figure}

\textbf{Orbital and spin Hall effect:}
In order to compute the OHC/SHC, we first construct a tight-binding (TB) model based on the parameters obtained from DFT, followed by Wannierization. To simplify the calculation, we consider bulk GeTe, as a 30 unit cell thick film. The band structure obtained from both methods is compared in Sec. S4 of the supplemental material. It shows a clear match between the DFT and TB bandstructures. 
The OHC and SHC are then computed using the Kubo formula,
\begin{equation}\label{OHC} 
  \sigma^{\gamma,\rm orb/spin}_{\alpha \beta}   =  -\frac{e} { N_k V_c} \sum_{ \vec  k} \Omega^{\gamma,\rm orb/spin}_{\alpha \beta} ({\vec  k}),
\end{equation}
where $\Omega^{\gamma,\rm orb/spin}_{\alpha \beta} ({\vec  k})
= \sum_n^{\rm occ} \Omega^{\gamma,\rm orb/spin}_{n,\alpha \beta} ({\vec  k})$ is the sum over occupied bands $n$ at a specific momentum point $\vec k$. We consider a dense $k$-mesh grid of $n_k =120^3$ for the summation. Here, $\alpha,\beta$, and $\gamma$ can be $x,y$ or $z$ such that $\alpha$ is the direction of the orbital/spin current, $\beta$ is the applied electric fied direction, whereas the orbital and spin angular momentum points towards $\gamma$.
The orbital Berry curvature for the Bloch state $u_{n\vec k}$ is computed using the relation
\begin{equation} \label{obc}         
 \Omega^{\gamma,\rm orb}_{n,\alpha\beta} ({\vec  k}) = 2 \hbar   \sum_{n^\prime \neq n} \frac {{\rm Im}[ \langle u_{n{\vec  k}} | \mathcal{J}^{\gamma,\rm orb}_\alpha | u_{n^\prime{\vec  k}} \rangle  
\langle u_{n^\prime{\vec  k}} | v_\beta | u_{n{\vec  k}} \rangle]} 
{(\varepsilon_{n^\prime \vec k}-\varepsilon_{n  \vec k} )^2}.
\end{equation}
Here, $v_{\alpha} =  \frac{1}{\hbar} \frac{\partial {\cal H} }{ \partial k_\alpha}$ is the velocity operator, and $\mathcal{J}^{\gamma,\rm orb}_\alpha = \frac{1}{2} \{v_\alpha, L_\gamma \}$ is the orbital current operator  
with $L_\gamma$ being the orbital angular momentum. For the spin Berry curvature, the orbtial current operator is replaced by the spin current operator $\mathcal{J}^{\gamma,\rm spin}_\alpha = \frac{1}{2} \{v_\alpha, S_\gamma \}$ in Eq. (\ref{obc}). 

Symmetry of the system dictates that there are 4 non-vanishing independent components of the OHC and SHC tensors~\cite{roy2022unconventional}, i.e., $\sigma^x_{zy},\sigma^y_{xz},$ $\sigma^z_{yx}$, and $\sigma^y_{yx}$. Other non-zero terms are related as $\sigma^{x}_{yz} = - \sigma^x_{zy}$, $\sigma^y_{zx} = - \sigma^y_{xz}$, and $\sigma^z_{xy} = -\sigma^z_{yx}$. 
The OHC and SHC as a function of the  Fermi energy $\mu$ are plotted in Fig.~\ref{fig:theory_combined} (a) and (b) respectively. For bulk GeTe, $E_F$ is set to 0, and we notice that at this energy level, the SHC is quite negligible, which is in agreement with the previous report~\cite{wang2020spin}. Interestingly, it can also be observed that as one deviates from $\mu = 0$, the magnitude of SHC rapidly rises. As mentioned earlier, there is a 0.25$e$ transfer from GeTe to the Py overlayer, which lowers the Fermi energy by 0.217 eV ($E_F^{het}$=-0.217 eV). The values of the conductivities at $E_F^{het}$ are listed in Table.~\ref{tab1}. 
\begin{table}[H]
    \centering
    \begin{tabular}{|c|c|c|c|c|}
    \hline
         \rowcolor{orange!50!white!40}Component & OHC &Eff. OHC ($=C_{FM}\times OHC$)& SHC & Total  \\
         \hline
         $\sigma^{x}_{zy}$&0.3327 &0.0127 &0.0011 & 0.0138\\
         $\sigma^{y}_{xz}$&0.1638 &0.0063 &0.0010 & 0.0073\\
         $\sigma^{z}_{yx}$ &-1.1672& -0.0446&0.2117 &0.1671\\
         $\sigma^{y}_{yx}$ & 0.3492 &0.0133 &0.0026 &0.0159\\
         \hline\hline
    \end{tabular}
    \caption{Independent tensor components of the OHC and SHC in GeTe with hole doping introduced by the Py. The conductivities are in the units $10^5\ \hbar/(2e)\ \Omega^{-1}m^{-1}$. We have considered the conversion factor ($C_{FM}$) to be 3.824\% for Py, based on the estimations made for Ni and Fe\cite{lee2021orbital} to calculate the effective OHC. Furthermore, we introduce the
earlier estimated charge transfer of 0.25e, which makes a Fermi level shift of -0.217 eV.}
    \label{tab1}
\end{table}

Our results indicate that the OHC is an order of magnitude larger than the SHC for the bulk GeTe. Such a large OHC is expected, since the system has broken inversion symmetry\cite{pratik}. However, we would also like to emphasize that, while the spin Hall current directly produces SOT, the orbital current must convert into spin current inside the Py to produce the SOT\cite{lee2021orbital} i.e. $\sigma_{tot} = \textup{SHC} + C_{FM}\times \textup{OHC}$, where $C_{FM}$ is the orbital to spin conversion factor for the Py. In our case, we take $C_{FM} = 3.824\%$ based on the estimations made for Ni and Fe\cite{lee2021orbital}. Hence, in combination with the charge transfer, GeTe produces a significant orbital and spin current, which contribute to the damping-like torque seen in the experiment.

In this study, we utilized polycrystalline GeTe films. In Table~\ref{tab1}, we calculate the SHC and OHC for the GeTe/Py system while considering only the single crystal nature of GeTe. The damping-like torque conductivity we measured is directly related to the SHC and OHC tensor, specifically to its off-diagonal elements shown in Table~\ref{tab1}. Although we use polycrystalline GeTe, each grain of a polycrystalline material exhibits local crystalline order with the z-axis being perpendicular to the interface. Therefore, we expect a large damping-like torque conductivity due to the averaging over all possible crystal orientations in perfect agreement with our experimental results. The presence of an unconventional damping-like torque $\sigma_{\rm DL}^{z}$  in experiment further support our argument. 
The study conducted by Jeon \textit{et al. }~\cite{jeon2021field} focused on epitaxial GeTe. Table~\ref{tab1} shows that in the epitaxial system, some components of the SHC and OHC tensor (e.g., \(\sigma^{y}_{xz}\)) can be negligible. This may be one of the reasons why they did not observe a large damping-like torque, potentially due to the specific crystallographic configuration in their work.

\textbf{Interfacial orbital Rashba effect:} 
The ORE arises in systems with broken inversion symmetry, which creates a chiral orbital texture in the momentum space. This is analogous to the well-known spin Rashba effect (SRE), where a spin texture is observed instead. While a strong SOC is needed for SRE, ORE is independent of the SOC, which makes the latter a more fundamental phenomenon. One important fingerprint of the SRE is the linear split of the bands with up and down spin characters. However, the orbital characters of both bands remain the same. As a result, if the bands are partially occupied, the Fermi surface comprises of two concentric circles with chiral spin and orbital textures having opposite spin chiralities, and same orbital chiralities. When an external electric field is applied, the Fermi surface shifts, creating an imbalance of the spin/orbital angular momentum, further leading to a net accumulation of the moments. This is known as the Edelstein effect~\cite{edelstein1990spin}.

\section{Discussion}

For systems like GeTe, where a strong SOC is present, both SRE and ORE are expected. While the SRE in GeTe has been reported earlier\cite{di2013electric}, the ORE is not presented in the literature. The ORE can be another contributor to the damping-like torque~\cite{johansson2021spin,chen2018giant,krishnia2024quantifying} in the experiments. We discuss both ORE and SRE in bulk GeTe and the possible Edelstein effect, keeping the overlayer in mind. We study the Rashba effect for 4 layers of GeTe. While the Rashba split has been reported near the $Z$ point for the bulk case \cite{di2013electric}, this high symmetry point is absent for the layered system. Rather, we find the Rashba split of the bands near the $\Gamma$ point, which is shown in Fig.~\ref{fig:theory_combined}(c) and (d). We estimate the Rashba constant from the energy and momentum shifts, which are $\Delta E_R = 0.0175 eV$ and $\Delta k_R = 0.0044$ \AA$^{-1}$ respectively. This corresponds to a Rashba constant of $\alpha_R \simeq 3.97$ eV\AA. In order to study the Rashba effect in detail, we computed the orbital and spin textures around the $\Gamma$ point using the eigenstates and energies obtained from the DFT based TB model. The orbital and spin texture are shown by Fig.~\ref{fig:theory_combined}(e) and (f), respectively. 

It shows that both orbital and spin textures are present and are also chiral. We 
also observe that while the spin textures for the inner and outer Fermi contours are opposite of each other, whereas they both have same chirality for the orbital moments. This is because the orbital Rashba is independent of the SOC and hence, both the bands have the same orbital characters.

We would like to emphasize that the Rashba moments can only give rise to SOT, if an Edelstein effect is possible, i.e., the Rashba bands must contribute to the Fermi contours. In the absence of the overlayer the bands giving rise to the Rashba split lies 0.5 eV below the Fermi level. However, we observe that in the presence of an overlayer (e.g.  Py, Ni), the interfacial interactions lead to a change in the band structure, where the bands at $\Gamma$ moves up in energy and lie at the Fermi. For demonstration in Fig.~\ref{fig:theory_combined}(g), we show the band structure of a Ni overlayered GeTe. The case of Py overlayer is not shown as the large number of bands in this case affects the clarity. 

Additionally, the hole doping due to the overlayer, as discussed in the previous section, makes the bands partially unoccupied, which leads to Fermi contours near the $\Gamma$ valley. As a result, the ORE and SRE give rise to Edelstein effect when electric field is applied. These accumulated orbital and spin moments, along with the OHC and SHC contribute to the SOT in the experiment. ORE and SRE in the presence of Edelstein effect can also produce the unconventional field-like torque $\sigma_{\rm FL}^{z}$, observed in the experiment.

In summary, our investigation into the heterostructure of GeTe/Py has successfully overcome a key limitation in vdW-based spintronics. We report a giant damping-like torque conductivity $\sigma{\rm_{DL}^{y}}$ of $(-1.25 \pm 0.11)\times 10^{5}~\hbar/ 2e~\Omega^{-1}$m$^{-1}$, which is the highest recorded for any FM/vdW interface and is comparable to benchmark heavy-metal systems. First-principles calculations elucidated the physical mechanism behind this remarkable efficiency, identifying it as a cooperative interplay between the spin Hall Effect, orbital Hall Effect, and orbital Rashba Effect, all enhanced by interfacial charge transfer. The work underscores the practical potential of vdW materials to facilitate efficient, low-power, and scalable electrical control of magnetization, paving the way for the commercial application of next-generation spintronic memory and logic devices.


\section{METHODS}



\subsection{Device fabrication}

Polycrystalline GeTe thin films were synthesized on Si(111) substrates using the pulsed laser deposition technique. A commercially available GeTe target was employed for the ablation process, which was carried out with a KrF excimer laser operating at 248 nm. The laser operated at a repetition rate of 5 Hz and delivered a fluence of 1.1 J cm$^{-2}$. The deposition took place in a pure Ar gas environment at an operating pressure of 1 $\times 10^{-4}$ mBar and a temperature of 250 $^\circ$C. The distance between the target and the substrate was maintained at 5 cm. After the deposition, the substrate was allowed to cool to room temperature, and a 3 nm Al layer was deposited to cap the GeTe films. This capping layer was added to prevent potential oxidation and chemical contamination when the films were exposed to the ambient atmosphere. Subsequently, the GeTe films were transferred to an ultra-high vacuum magnetron sputtering chamber (AJA Orion 8) for the deposition of Py(8 nm)/AlO$_x$(2 nm). The deposition of both Py and Al was carried out at a fixed working pressure of 2$\times 10^{-3}$~Torr, with the base pressure of the sputtering chamber maintained below 5$\times 10^{-8}$~Torr. The growth rates for the Al and Py layers were 0.20~$\mathring{\mathrm{A}}/ \rm s$ and 0.23~$\mathring{\mathrm{A}}/ \rm s$, respectively. The growth conditions of Py and Al was carefully choose following our earlier studies~\cite{bangar2022large,bangar2023optimization,bansal2019extrinsic} to avoid damage to GeTe.
The GeTe/Py microstrip devices were fabricated using a mask aligner (MA/BA6, Suss Microtech Lithography) and etched using Ar-ion etching inside our AJA sputter chamber through substrate bias. Subsequently, Cr/Au contact pads were patterned using a mask aligner and liftoff technique.

\subsection{Thin film characterization}

The structural analysis of the GeTe thin films was conducted using a PANalytical X-ray diffractometer equipped with a Cu K$_{\alpha}$ source ($\lambda$ = 1.5406 \AA). Raman measurements utilized a Renishaw inVia confocal microscope with a 532 nm laser wavelength in backscattering geometry and 2400 lines per mm grating, delivering 5 mW of power. To assess surface topography and roughness, atomic force microscopy scans were carried out using an Asylum Research MFP-3D system operating in tapping mode. Asylum Research Probes (AC240TS-R3) cantilevers were employed for these scans. The surface chemistry of the GeTe thin films was investigated via XPS measurements conducted on a Kratos Analytical Ltd. AXIS Supra system.

\subsection{SOT investigation}

For the investigation of SOTs, we fabricated microstrip devices measuring 80 $\mu$m in length and 20 $\mu $m in width for the STFMR measurements. These devices were designed to apply an RF current and simultaneously measure the resulting DC voltage using a bias-tee [Fig.\ref{fig:2}(a)]. We supplied RF power of +7 dBm using an R\&S signal generator (SMB 100A). An amplitude modulation-based method was employed to determine the rectified DC voltage\cite{kumar2021large}. To manipulate the magnetic field angle $\varphi$, we used a GMW 5201 projected field magnet and swept $\varphi$ from 0 to 360 degrees.  

\subsection{DFT calculation}
The technical aspects of the theoretical investigation concerning interfacial charge transfer in the heterostructure and the calculation of OHC/SHC of the bulk system are discussed below.\\
\noindent\textit{(a) Heterostructure}: The DFT results for the heterostructure were obtained using the Vienna Ab initio Simulation Package (VASP)~\cite{kresse1996efficient}. The self-consistent field (SCF) calculation was performed using GGA pseudopotentials. The convergence was achieved in a $k-$mesh of $12\times\ 10\times 1$, with  a plane wave energy cut-off of 200 Ry.\\
\noindent\textit{(b) Transport properties}: Bulk GeTe was considered to compute the OHC and SHC. First, we carried out an SCF calculation using the Quantum Espresso package~\cite{giannozzi2009quantum}, in a k-mesh of $12\times 12\times 10$ with norm-conserving Perdew, Burke, and Ernzerhof (PBE) exchange-correlation functionals~\cite{perdew1996generalized}. This was followed by a non self-consistent field (NSCF) calculation considering a larger k-mesh ($20 \times 20 \times 16$). Following this, a TB model Hamiltonian was constructed using the hopping parameters extracted from DFT via the Wannier90 code~\cite{pizzi2020wannier90}. Finally, OHC and SHC were computed using the Kubo formula (see Eq. \ref{OHC}) on a dense $120\times 120\times 120$  k-mesh.
\begin{acknowledgement}
The partial support from the Ministry of Human Resource Development under the IMPRINT program (Grants Nos. 7519 and 7058), the Ministry of Electronics and Information Technology (MeitY) (Approval No: $Y-20/8/2024-R\&D-E$), the Science and Engineering Research Board (SERB File No. CRG/2018/001012 and CRG/2022/002821), Joint Advanced Technology Centre at IIT Delhi, Grand Challenge Project, IIT Delhi are gratefully acknowledged. Financial support from Swedish Research Council, Distinguished professor (Grant: 2024-01943) and Horizon 2020 research and innovation programme (Grant No. 835068 “TOPSPIN”) are also acknowledged. The authors also acknowledged the Central Research Facility, IIT Delhi, for providing facilities for sample characterization. We also acknowledge nanocsale research facility, IIT Delhi for fabrication facilities and the Department of Physics, IIT Delhi for the pulsed laser deposition system. H.B. gratefully acknowledges the financial support from the Council of Scientific and Industrial Research (CSIR), Government of India. 

\end{acknowledgement}

\bibliography{2_bibliography}

@STRING{apl = {Appl.\ Phys.\ Lett.}}

@STRING{ntc = {Nat.\ Commun.}}

@STRING{ntp = {Nat.\ Phys.}}

@STRING{prb = {Phys.\ Rev.\ B}}

@STRING{prm = {Phys.\ Rev.\ Mater.}}

@STRING{sc = {Science}}

@article{hirsch1999spin,
  title = "{Spin Hall Effect}",
  author = {Hirsch, J. E.},
  journal = {Phys. Rev. Lett.},
  volume = {83},
  issue = {9},
  pages = {1834--1837},
  numpages = {0},
  year = {1999},
  month = {Aug},
  publisher = {American Physical Society},
  doi = {10.1103/PhysRevLett.83.1834},
  url = {https://link.aps.org/doi/10.1103/PhysRevLett.83.1834}
}

@article{park2011orbital,
  title={Orbital-angular-momentum based origin of Rashba-type surface band splitting},
  author={Park, Seung Ryong and Kim, Choong H and Yu, Jaejun and Han, Jung Hoon and Kim, Changyoung},
  journal={Phys. Rev. Lett.},
  volume={107},
  number={15},
  pages={156803},
  year={2011},
  publisher={APS}
}

@article{choi2023observation,
  title={Observation of the orbital Hall effect in a light metal Ti},
  author={Choi, Young-Gwan and Jo, Daegeun and Ko, Kyung-Hun and Go, Dongwook and Kim, Kyung-Han and Park, Hee Gyum and Kim, Changyoung and Min, Byoung-Chul and Choi, Gyung-Min and Lee, Hyun-Woo},
  journal={Nature},
  volume={619},
  number={7968},
  pages={52--56},
  year={2023},
  publisher={Nature Publishing Group UK London}
}

@article{dieny2020opportunities,
  title={Opportunities and challenges for spintronics in the microelectronics industry},
  author={Dieny, Bernard and Prejbeanu, Ioan Lucian and Garello, Kevin and Gambardella, Pietro and Freitas, Paulo and Lehndorff, Ronald and Raberg, Wolfgang and Ebels, Ursula and Demokritov, Sergej O and Akerman, Johan and others},
  journal={Nat. Electron.},
  volume={3},
  number={8},
  pages={446--459},
  year={2020},
  publisher={Nature Publishing Group UK London}
}

@article{kumar2025spin,
  title={Spin-wave-mediated mutual synchronization and phase tuning in spin Hall nano-oscillators},
  author={Kumar, Akash and Chaurasiya, Avinash Kumar and Gonz{\'a}lez, Victor H and Behera, Nilamani and Alem{\'a}n, Ademir and Khymyn, Roman and Awad, Ahmad A and {\AA}kerman, Johan},
  journal=ntp,
  volume={21},
  number={2},
  pages={245--252},
  year={2025},
  publisher={Nature Publishing Group UK London}
}

@article{bangar2023optimization,
  title={Optimization of growth of large-area SnS thin films and heterostructures for spin pumping and spin-orbit torque},
  author={Bangar, Himanshu and Gupta, Pankhuri and Singh, Rajendra and Muduli, Pranaba Kishor and Dewan, Sheetal and Das, Samaresh},
  journal=prm,
  volume={7},
  number={9},
  pages={094406},
  year={2023},
  publisher={APS}
}

@article{shelimova1993crystal,
  title={Crystal structure, phase transitions, and mechanical properties of GeTe-based solid solutions in the GeTe-PbTe-MTe systems (M= Mn, Sc, La)},
  author={Shelimova, LE and Karpinskii, OG and Avilov, ES and Kretova, MA},
  journal={Inorg. Mater.},
  volume={29},
  number={11},
  pages={1291--1298},
  year={1993},
  publisher={New York: Consultants Bureau, 1965-}
}

@article{yang2024harnessing,
  title={Harnessing synergy of spin and orbital currents in heavy metal/ferromagnet multilayers},
  author={Yang, Yumin and Xie, Zhicheng and Zhao, Zhiyuan and Lei, Na and Zhao, Jianhua and Wei, Dahai},
  journal={Commun. Phys.},
  volume={7},
  number={1},
  pages={336},
  year={2024},
  publisher={Nature Publishing Group UK London}
}

@article{an2023electrical,
  title={Electrical Manipulation of Orbital Current Via Oxygen Migration in Ni$_{81}$Fe$_{19}$/CuO$_{\rm x}$/TaN Heterostructure},
  author={An, Taiyu and Cui, Bin and Zhang, Mingfang and Liu, Fufu and Cheng, Shaobo and Zhang, Kuikui and Ren, Xue and Liu, Liang and Cheng, Bin and Jiang, Changjun and others},
  journal={Adv. Mater.},
  volume={35},
  number={25},
  pages={2300858},
  year={2023},
  publisher={Wiley Online Library}
}

@article{hirohata2020review,
  title={Review on spintronics: Principles and device applications},
  author={Hirohata, Atsufumi and Yamada, Keisuke and Nakatani, Yoshinobu and Prejbeanu, Ioan-Lucian and Di{\'e}ny, Bernard and Pirro, Philipp and Hillebrands, Burkard},
  journal={J. Magn. Magn. Mater.},
  volume={509},
  pages={166711},
  year={2020},
  publisher={Elsevier}
}

@article{krishnia2024quantifying,
  title={Quantifying the large contribution from orbital Rashba--Edelstein effect to the effective damping-like torque on magnetization},
  author={Krishnia, S and Bony, B and Rongione, E and Vicente-Arche, L Moreno and Denneulin, T and Pezo, A and Lu, Y and Dunin-Borkowski, RE and Collin, S and Fert, A and others},
  journal={APL Mater.},
  volume={12},
  number={5},
  year={2024},
  publisher={AIP Publishing}
}

@article{chen2018giant,
  title={Giant antidamping orbital torque originating from the orbital Rashba-Edelstein effect in ferromagnetic heterostructures},
  author={Chen, Xi and Liu, Yang and Yang, Guang and Shi, Hui and Hu, Chen and Li, Minghua and Zeng, Haibo},
  journal=ntc,
  volume={9},
  number={1},
  pages={2569},
  year={2018},
  publisher={Nature Publishing Group UK London}
}

@article{johansson2021spin,
  title={Spin and orbital Edelstein effects in a two-dimensional electron gas: Theory and application to SrTiO$_3$ interfaces},
  author={Johansson, Annika and G{\"o}bel, B{\"o}rge and Henk, J{\"u}rgen and Bibes, Manuel and Mertig, Ingrid},
  journal={Phys. Rev. Res.},
  volume={3},
  number={1},
  pages={013275},
  year={2021},
  publisher={APS}
}

@article{cysne2025orbitronics,
  title={Orbitronics in two-dimensional materials},
  author={Cysne, Tarik P and Canonico, Luis M and Costa, Marcio and Muniz, RB and Rappoport, Tatiana G},
  journal={npj Spintronics},
  volume={3},
  number={1},
  pages={39},
  year={2025},
  publisher={Nature Publishing Group UK London}
}

@article{fons2010phase,
  title={Phase transition in crystalline GeTe: Pitfalls of averaging effects},
  author={Fons, Paul and Kolobov, Alexander V and Krbal, Milos and Tominaga, Junji and Andrikopoulos, KS and Yannopoulos, SN and Voyiatzis, GA and Uruga, T},
  journal={Phys. Rev. B},
  volume={82},
  number={15},
  pages={155209},
  year={2010},
  publisher={APS}
}

@article{di2012electric,
  title={Electric control of the giant Rashba effect in bulk GeTe.},
  author={Di Sante, Domenico and Barone, Paolo and Bertacco, Riccardo and Picozzi, Silvia},
  journal={Adv. Mater},
  volume={25},
  number={4},
  pages={509--513},
  year={2013}
}

@article{wang2020spin,
  title={Spin Hall effect in prototype Rashba ferroelectrics GeTe and SnTe},
  author={Wang, Haihang and Gopal, Priya and Picozzi, Silvia and Curtarolo, Stefano and Buongiorno Nardelli, Marco and S{\l}awi{\'n}ska, Jagoda},
  journal={Npj Comput. Mater.},
  volume={6},
  number={1},
  pages={7},
  year={2020},
  publisher={Nature Publishing Group UK London}
}

@article{lee2021orbital,
  title={Orbital torque in magnetic bilayers},
  author={Lee, Dongjoon and Go, Dongwook and Park, Hyeon-Jong and Jeong, Wonmin and Ko, Hye-Won and Yun, Deokhyun and Jo, Daegeun and Lee, Soogil and Go, Gyungchoon and Oh, Jung Hyun and others},
  journal={Nat. Commun.},
  volume={12},
  number={1},
  pages={6710},
  year={2021},
  publisher={Nature Publishing Group UK London}
}

@article{roy2022unconventional,
  title={Unconventional spin Hall effects in nonmagnetic solids},
  author={Roy, Arunesh and Guimar{\~a}es, Marcos HD and S{\l}awi{\'n}ska, Jagoda},
  journal=prm,
  volume={6},
  number={4},
  pages={045004},
  year={2022},
  publisher={APS}
}

@article{sala2022giant,
  title={Giant orbital Hall effect and orbital-to-spin conversion in 3d, 5d, and 4f metallic heterostructures},
  author={Sala, Giacomo and Gambardella, Pietro},
  journal={Phys. Rev. Res.},
  volume={4},
  number={3},
  pages={033037},
  year={2022},
  publisher={APS}
}

@article{kontani2009giant,
  title={Giant orbital Hall effect in transition metals: Origin of large spin and anomalous Hall effects},
  author={Kontani, Hiroshi and Tanaka, T and Hirashima, DS and Yamada, K and Inoue, J},
  journal={Phys. Rev. Lett.},
  volume={102},
  number={1},
  pages={016601},
  year={2009},
  publisher={APS}
}

@article{go2018intrinsic,
  title={Intrinsic spin and orbital Hall effects from orbital texture},
  author={Go, Dongwook and Jo, Daegeun and Kim, Changyoung and Lee, Hyun-Woo},
  journal={Phys. Rev. Lett.},
  volume={121},
  number={8},
  pages={086602},
  year={2018},
  publisher={APS}
}

@article{jo2018gigantic,
  title={Gigantic intrinsic orbital Hall effects in weakly spin-orbit coupled metals},
  author={Jo, Daegeun and Go, Dongwook and Lee, Hyun-Woo},
  journal={Phys. Rev. B},
  volume={98},
  number={21},
  pages={214405},
  year={2018},
  publisher={APS}
}

@article{bansal2019extrinsic,
  title="{Extrinsic Spin-Orbit Coupling Induced Enhanced Spin Pumping in Few-Layer MoS$_2$/Py}",
  author={Bansal, Rajni and Kumar, Akash and Chowdhury, Niru and Sisodia, Naveen and Barvat, Arun and Dogra, Anjana and Pal, Prabir and Muduli, PK},
  journal={J. Magn. Magn. Mater.},
  volume={476},
  pages={337--341},
  year={2019},
  publisher={Elsevier}
}

@article{agarwal2023strong,
  title={Strong impact of crystalline twins on the amplitude and azimuthal dependence of THz emission from epitaxial NiO/Pt},
  author={Agarwal, Rekha and Kumar, Sandeep and Chowdhury, Niru and Khan, Kacho Imtiyaz Ali and Yadav, Ekta and Kumar, Sunil and Muduli, PK},
  journal={Appl. Phys. Lett.},
  volume={122},
  number={8},
  year={2023},
  publisher={AIP Publishing}
}

@article{farle1998ferromagnetic,
  title={Ferromagnetic resonance of ultrathin metallic layers},
  author={Farle, Michael},
  journal={Rep. Prog. Phys.},
  volume={61},
  number={7},
  pages={755},
  year={1998},
  publisher={IOP Publishing}
}

@article{bychkov1984properties,
  title="{Properties of a 2D Electron Gas with Lifted Spectral Degeneracy}",
  author={Bychkov, Yua A and Rashba, {\'E} I},
  journal={JETP Lett.},
  volume={39},
  number={2},
  pages={78--81},
  year={1984}
}

@article{edelstein1990spin,
  title={Spin Polarization of Conduction Electrons Induced by Electric Current in Two-Dimensional Asymmetric Electron Systems},
  author={Edelstein, Victor M},
  journal={Solid State Commun.},
  volume={73},
  number={3},
  pages={233--235},
  year={1990},
  publisher={Elsevier}
}

@article{kao2022deterministic,
  title={Deterministic switching of a perpendicularly polarized magnet using unconventional spin--orbit torques in WTe$_2$},
  author={Kao, I-Hsuan and Muzzio, Ryan and Zhang, Hantao and Zhu, Menglin and Gobbo, Jacob and Yuan, Sean and Weber, Daniel and Rao, Rahul and Li, Jiahan and Edgar, James H and others},
  journal={Nat. Mater.},
  volume={21},
  number={9},
  pages={1029--1034},
  year={2022},
  publisher={Nature Publishing Group UK London}
}

@article{shin2022spin,
  title={Spin--Orbit Torque Switching in an All-Van der Waals Heterostructure},
  author={Shin, Inseob and Cho, Won Joon and An, Eun-Su and Park, Sungyu and Jeong, Hyeon-Woo and Jang, Seong and Baek, Woon Joong and Park, Seong Yong and Yang, Dong-Hwan and Seo, Jun Ho and others},
  journal={Adv. Mater.},
  volume={34},
  number={8},
  pages={2101730},
  year={2022},
  publisher={Wiley Online Library}
}

@article{bainsla2024large,
  title={Large out-of-plane spin--orbit torque in topological Weyl semimetal TaIrTe$_4$},
  author={Bainsla, Lakhan and Zhao, Bing and Behera, Nilamani and Hoque, Anamul Md and Sj{\"o}str{\"o}m, Lars and Martinelli, Anna and Abdel-Hafiez, Mahmoud and {\AA}kerman, Johan and Dash, Saroj P},
  journal={Nat. Commun.},
  volume={15},
  number={1},
  pages={4649},
  year={2024},
  publisher={Nature Publishing Group UK London}
}

@article{dyakonov1971current,
  title={Current-Induced Spin Orientation of Electrons in Semiconductors},
  author={Dyakonov, Mikhail I and Perel, VI},
  journal={Phys. Lett. A},
  volume={35},
  number={6},
  pages={459--460},
  year={1971},
  publisher={Elsevier}
}

@PREAMBLE{
 "\providecommand{\noopsort}[1]{}" 
 # "\providecommand{\singleletter}[1]{#1}%" 
}

@article{el2023observation,
  title={Observation of the orbital inverse Rashba--Edelstein effect},
  author={El Hamdi, Anas and Chauleau, Jean-Yves and Boselli, Margherita and Thibault, Cl{\'e}mentine and Gorini, Cosimo and Smogunov, Alexander and Barreteau, Cyrille and Gariglio, Stefano and Triscone, Jean-Marc and Viret, Michel},
  journal={Nat. Phys.},
  volume={19},
  number={12},
  pages={1855--1860},
  year={2023},
  publisher={Nature Publishing Group UK London}
}

@article{kalra2015role,
  title={The role of atomic vacancies on phonon confinement in $\alpha$-GeTe},
  author={Kalra, Geetanjali and Murugavel, Sevi},
  journal={AIP Adv.},
  volume={5},
  numb={4},
  year={2015},
  publisher={AIP Publishing}
}

@article{liebmann2016giant,
  title={Giant Rashba-type spin splitting in ferroelectric GeTe (111)},
  author={Liebmann, Marcus and Rinaldi, Christian and Di Sante, Domenico and Kellner, Jens and Pauly, Christian and Wang, Rui Ning and Boschker, Jos Emiel and Giussani, Alessandro and Bertoli, Stefano and Cantoni, Matteo and others},
  journal={Adv. Mater.},
  volume={28},
  number={3},
  pages={560--565},
  year={2016}
}

@article{myers1999current,
  title={Current-induced switching of domains in magnetic multilayer devices},
  author={Myers, EB and Ralph, DC and Katine, JA and Louie, RN and Buhrman, RA},
  journal={Science},
  volume={285},
  number={5429},
  pages={867--870},
  year={1999},
  publisher={American Association for the Advancement of Science}
}

@article{miron2011perpendicular,
  title={Perpendicular switching of a single ferromagnetic layer induced by in-plane current injection},
  author={Miron, Ioan Mihai and Garello, Kevin and Gaudin, Gilles and Zermatten, Pierre-Jean and Costache, Marius V and Auffret, St{\'e}phane and Bandiera, S{\'e}bastien and Rodmacq, Bernard and Schuhl, Alain and Gambardella, Pietro},
  journal={Nature},
  volume={476},
  number={7359},
  pages={189--193},
  year={2011},
  publisher={Nature Publishing Group UK London}
}

@article{fukami2016spin,
  title={A spin--orbit torque switching scheme with collinear magnetic easy axis and current configuration},
  author={Fukami, Shunsuke and Anekawa, T and Zhang, C and Ohno, H},
  journal={Nat. Nanotechnol.},
  volume={11},
  number={7},
  pages={621--625},
  year={2016},
  publisher={Nature Publishing Group UK London}
}

@article{mellnik2014spin,
  title={Spin-transfer torque generated by a topological insulator},
  author={Mellnik, AR and Lee, JS and Richardella, A and Grab, JL and Mintun, PJ and Fischer, Mark H and Vaezi, Abolhassan and Manchon, Aurelien and Kim, E A and Samarth, Nitin and Ralph, D C},
  journal={Nature},
  volume={511},
  numb={7510},
  pages={449--451},
  year={2014},
  publisher={Nature Publishing Group UK London},
  url={https://www.nature.com/articles/nature13534},
}

@article{di2013electric,
  title={Electric control of the giant Rashba effect in bulk GeTe},
  author={Di Sante, Domenico and Barone, Paolo and Bertacco, Riccardo and Picozzi, Silvia},
  journal={Adv. Mater.},
  volume={25},
  number={4},
  pages={509--513},
  year={2013},
  publisher={Wiley Online Library}
}

@article{safeer2019room,
  title={Room-temperature spin Hall effect in graphene/MoS$_2$ van der Waals heterostructures},
  author={Safeer, CK and Ingla-Ayn{\'e}s, Josep and Herling, Franz and Garcia, Jos{\'e} H and Vila, Marc and Ontoso, Nerea and Calvo, M Reyes and Roche, Stephan and Hueso, Luis E and Casanova, F{\`e}lix},
  journal={Nano Lett.},
  volume={19},
  number={2},
  pages={1074--1082},
  year={2019},
  publisher={ACS Publications}
}

@article{pizzi2020wannier90,
  title={Wannier90 as a community code: new features and applications},
  author={Pizzi, Giovanni and Vitale, Valerio and Arita, Ryotaro and Bl{\"u}gel, Stefan and Freimuth, Frank and G{\'e}ranton, Guillaume and Gibertini, Marco and Gresch, Dominik and Johnson, Charles and Koretsune, Takashi and others},
  journal={J. Phys.: Condens. Matter},
  volume={32},
  number={16},
  pages={165902},
  year={2020},
  publisher={IOP Publishing}
}

@article{perdew1996generalized,
  title={Generalized gradient approximation made simple},
  author={Perdew, John P and Burke, Kieron and Ernzerhof, Matthias},
  journal={Phys. Rev. Lett.},
  volume={77},
  number={18},
  pages={3865},
  year={1996},
  publisher={APS}
}

@article{giannozzi2009quantum,
  title={QUANTUM ESPRESSO: a modular and open-source software project for quantumsimulations of materials},
  author={Giannozzi, Paolo and Baroni, Stefano and Bonini, Nicola and Calandra, Matteo and Car, Roberto and Cavazzoni, Carlo and Ceresoli, Davide and Chiarotti, Guido L and Cococcioni, Matteo and Dabo, Ismaila and others},
  journal={J. Phys.: Condens. Matter},
  volume={21},
  number={39},
  pages={395502},
  year={2009},
  publisher={IOP Publishing}
}

@article{kresse1996efficient,
  title={Efficient iterative schemes for ab initio total-energy calculations using a plane-wave basis set},
  author={Kresse, Georg and Furthm{\"u}ller, J{\"u}rgen},
  journal={Phys. Rev. B},
  volume={54},
  number={16},
  pages={11169},
  year={1996},
  publisher={APS}
}

@article{shi2019all,
  title={All-electric magnetization switching and Dzyaloshinskii--Moriya interaction in WTe$_2$/ferromagnet heterostructures},
  author={Shi, Shuyuan and Liang, Shiheng and Zhu, Zhifeng and Cai, Kaiming and Pollard, Shawn D and Wang, Yi and Wang, Junyong and Wang, Qisheng and He, Pan and Yu, Jiawei and others},
  journal={Nat. Nanotechnol.},
  volume={14},
  number={10},
  pages={945--949},
  year={2019},
  publisher={Nature Publishing Group UK London}
}

@article{liu2023field,
  title={Field-free switching of perpendicular magnetization at room temperature using out-of-plane spins from TaIrTe4},
  author={Liu, Yakun and Shi, Guoyi and Kumar, Dushyant and Kim, Taeheon and Shi, Shuyuan and Yang, Dongsheng and Zhang, Jiantian and Zhang, Chenhui and Wang, Fei and Yang, Shuhan and others},
  journal={Nat. Electron.},
  volume={6},
  number={10},
  pages={732--738},
  year={2023},
  publisher={Nature Publishing Group UK London}
}

@article{xu2020high,
  title={High spin hall conductivity in large-area type-II Dirac semimetal PtTe$_2$},
  author={Xu, Hongjun and Wei, Jinwu and Zhou, Hengan and Feng, Jiafeng and Xu, Teng and Du, Haifeng and He, Congli and Huang, Yuan and Zhang, Junwei and Liu, Yizhou and others},
  journal={Adv. 
 Mater.},
  volume={32},
  number={17},
  pages={2000513},
  year={2020},
  publisher={Wiley Online Library}
}

@article{jeon2021field,
  title={Field-like spin--orbit torque induced by bulk Rashba channels in GeTe/NiFe bilayers},
  author={Jeon, Jeehoon and Cho, Seong Won and Lee, OukJae and Hong, Jinki and Kwak, Joon Young and Han, Seungwu and Jung, Soonho and Kim, Yunseok and Ko, Hye-Won and Lee, Suyoun and others},
  journal={NPG Asia Mater.},
  volume={13},
  number={1},
  pages={76},
  year={2021},
  publisher={Nature Publishing Group UK London}
}

@article{krempasky2016disentangling,
  title={Disentangling bulk and surface Rashba effects in ferroelectric $\alpha$-GeTe},
  author={Krempask{\`y}, J{\'u}lius and Volfov{\'a}, Henrieta and Muff, Stefan and Pilet, Nicolas and Landolt, Gabriel and Radovi{\'c}, Miroslav and Shi, Ming and Kriegner, Dominik and Hol{\`y}, V{\'a}clav and Braun, J and others},
  journal={Phys. Rev. B},
  volume={94},
  number={20},
  pages={205111},
  year={2016},
  publisher={APS}
}

@article{wang2016ordered,
  title={Ordered Peierls distortion prevented at growth onset of GeTe ultra-thin films},
  author={Wang, Ruining and Campi, Davide and Bernasconi, Marco and Momand, Jamo and Kooi, Bart J and Verheijen, Marcel A and Wuttig, Matthias and Calarco, Raffaella},
  journal={Sci. Rep.},
  volume={6},
  number={1},
  pages={32895},
  year={2016},
  publisher={Nature Publishing Group UK London}
}

@article{zhang2016research,
  title={Research Update: Spin transfer torques in permalloy on monolayer MoS$_2$},
  author={Zhang, Wei and Sklenar, Joseph and Hsu, Bo and Jiang, Wanjun and Jungfleisch, Matthias B and Xiao, Jiao and Fradin, Frank Y and Liu, Yaohua and Pearson, John E and Ketterson, John B and others},
  journal={APL Mater.},
  volume={4},
  number={3},
  pages={032302},
  year={2016},
  publisher={AIP Publishing LLC}
}

@article{macneill2017thickness,
  title={Thickness dependence of spin-orbit torques generated by WTe$_2$},
  author={MacNeill, David and Stiehl, Gregory M and Guimaraes, Marcos HD and Reynolds, Neal D and Buhrman, Robert A and Ralph, Daniel C},
  journal=prb,
  volume={96},
  number={5},
  pages={054450},
  year={2017},
  publisher={APS}
}

@article{guimaraes2018spin,
  title={Spin--orbit torques in NbSe$_2$/permalloy bilayers},
  author={Guimaraes, Marcos HD and Stiehl, Gregory M and MacNeill, David and Reynolds, Neal D and Ralph, Daniel C},
  journal={Nano Lett.},
  volume={18},
  number={2},
  pages={1311--1316},
  year={2018},
  publisher={ACS Publications}
}

@article{kurebayashi2022magnetism,
  title={Magnetism, symmetry and spin transport in van der Waals layered systems},
  author={Kurebayashi, Hidekazu and Garcia, Jose H and Khan, Safe and Sinova, Jairo and Roche, Stephan},
  journal={Nat. Rev. Phys.},
  volume={4},
  number={3},
  pages={150--166},
  year={2022},
  publisher={Nature Publishing Group UK London}
}

@article{stiehl2019layer,
  title={Layer-dependent spin-orbit torques generated by the centrosymmetric transition metal dichalcogenide $\beta$-MoTe$_2$},
  author={Stiehl, Gregory M and Li, Ruofan and Gupta, Vishakha and El Baggari, Ismail and Jiang, Shengwei and Xie, Hongchao and Kourkoutis, Lena F and Mak, Kin Fai and Shan, Jie and Buhrman, Robert A and others},
  journal=prb,
  volume={100},
  number={18},
  pages={184402},
  year={2019},
  publisher={APS}
}

@article{bangar2023large,
  title={Large Spin Hall Conductivity in Epitaxial Thin Films of Kagome Antiferromagnet Mn$_3$Sn at Room Temperature},
  author={Bangar, Himanshu and Khan, Kacho Imtiyaz Ali and Kumar, Akash and Chowdhury, Niru and Muduli, Prasanta Kumar and Muduli, Pranaba Kishor},
  journal={Adv. Quantum Technol.},
  volume={6},
  number={1},
  pages={2200115},
  year={2023},
  publisher={Wiley Online Library}
}

@article{liu2012spin,
  title={Spin-torque switching with the giant spin Hall effect of tantalum},
  author={Liu, Luqiao and Pai, Chi-Feng and Li, Y and Tseng, HW and Ralph, DC and Buhrman, RA},
  journal={Science},
  volume={336},
  number={6081},
  pages={555--558},
  year={2012},
  publisher={American Association for the Advancement of Science}
}

@article{sui2017giant,
  title={Giant enhancement of the intrinsic spin Hall conductivity in $\beta$-tungsten via substitutional doping},
  author={Sui, Xuelei and Wang, Chong and Kim, Jinwoong and Wang, Jianfeng and Rhim, SH and Duan, Wenhui and Kioussis, Nicholas},
  journal={Phys. Rev. B},
  volume={96},
  number={24},
  pages={241105},
  year={2017},
  publisher={APS}
}

@Article{You2021,
author={You, Yunfeng
and Bai, Hua
and Feng, Xiaoyu
and Fan, Xiaolong
and Han, Lei
and Zhou, Xiaofeng
and Zhou, Yongjian
and Zhang, Ruiqi
and Chen, Tongjin
and Pan, Feng
and Song, Cheng},
title={Cluster magnetic octupole induced out-of-plane spin polarization in antiperovskite antiferromagnet},
journal=ntc,
year={2021},
month={Nov},
day={11},
volume={12},
number={1},
pages={6524},
issn={2041-1723},
doi={10.1038/s41467-021-26893-6},
url={https://doi.org/10.1038/s41467-021-26893-6}
}

@article{zhou2020magnetic,
  title="{Magnetic asymmetry induced anomalous spin-orbit torque in IrMn}",
  author={Zhou, Jing and Shu, Xinyu and Liu, Yaohua and Wang, Xiao and Lin, Weinan and Chen, Shaohai and Liu, Liang and Xie, Qidong and Hong, Tao and Yang, Ping and others},
  journal={Phys. Rev. B},
  volume={101},
  number={18},
  pages={184403},
  year={2020},
  publisher={APS}
}

@article{kittel1948theory,
  title={On the theory of ferromagnetic resonance absorption},
  author={Kittel, Charles},
  journal={Phys. Rev.},
  volume={73},
  number={2},
  pages={155},
  year={1948},
  publisher={APS}
}

@article{liu2020two,
  title="{Two-dimensional materials for energy-efficient spin--orbit torque devices}",
  author={Liu, Yuting and Shao, Qiming},
  journal={ACS Nano},
  volume={14},
  number={8},
  pages={9389--9407},
  year={2020},
  publisher={ACS Publications}
}

@article{de2007ferromagnetic,
  title={A ferromagnetic resonance study of NiFe alloy thin films},
  author={De Sihues, M Diaz and Durante-Rinc{\'o}n, CA and Fermin, JR},
  journal={J. Magn. Magn. Mater.},
  volume={316},
  number={2},
  pages={e462--e465},
  year={2007},
  publisher={Elsevier}
}

@article{macneill2017control,
  title="{Control of spin--orbit torques through crystal symmetry in WTe$_2$/ferromagnet bilayers}",
  author={MacNeill, D and Stiehl, GM and Guimaraes, MHD and Buhrman, RA and Park, J and Ralph, DC},
  journal={Nat. Phys.},
  volume={13},
  number={3},
  pages={300--305},
  year={2017},
  publisher={Nature Publishing Group}
}

@article{liu2021symmetry,
  title="{Symmetry-dependent field-free switching of perpendicular magnetization}",
  author={Liu, Liang and Zhou, Chenghang and Shu, Xinyu and Li, Changjian and Zhao, Tieyang and Lin, Weinan and Deng, Jinyu and Xie, Qidong and Chen, Shaohai and Zhou, Jing and others},
  journal={Nat. Nanotechnol.},
  volume={16},
  number={3},
  pages={277--282},
  year={2021},
  publisher={Nature Publishing Group}
}

@article{liu2019current,
  title="{Current-induced magnetization switching in all-oxide heterostructures}",
  author={Liu, Liang and Qin, Qing and Lin, Weinan and Li, Changjian and Xie, Qidong and He, Shikun and Shu, Xinyu and Zhou, Chenghang and Lim, Zhishiuh and Yu, Jihang and others},
  journal={Nat. Nanotechnol.},
  volume={14},
  number={10},
  pages={939--944},
  year={2019},
  publisher={Nature Publishing Group}
}

@article{shao2021roadmap,
  title={Roadmap of spin--orbit torques},
  author={Shao, Qiming and Li, Peng and Liu, Luqiao and Yang, Hyunsoo and Fukami, Shunsuke and Razavi, Armin and Wu, Hao and Wang, Kang and Freimuth, Frank and Mokrousov, Yuriy and others},
  journal={IEEE Trans. Magn.},
  volume={57},
  number={7},
  pages={1--39},
  year={2021},
  publisher={IEEE}
}

@article{rothschild2022generation,
  title={Generation of spin currents by the orbital Hall effect in Cu and Al and their measurement by a Ferris-wheel ferromagnetic resonance technique at the wafer level},
  author={Rothschild, Amit and Am-Shalom, Nadav and Bernstein, Nirel and Meron, Ma'yan and David, Tal and Assouline, Benjamin and Frohlich, Elichai and Xiao, Jiewen and Yan, Binghai and Capua, Amir},
  journal={Phys. Rev. B},
  volume={106},
  number={14},
  pages={144415},
  year={2022},
  publisher={APS}
}

@article{zhang2020tuning,
  title={Tuning spin Hall conductivity in GeTe by ferroelectric polarization},
  author={Zhang, Wenxu and Teng, Zhao and Zeng, Huizhong and Zhang, Hongbin and {\v{Z}}elezn{\`y}, Jakub and Zhang, Wanli},
  journal={Phys. Status Solidi B},
  volume={257},
  number={9},
  pages={2000143},
  year={2020},
  publisher={Wiley Online Library}
}

@article{chatterji2015anomalous,
  title={Anomalous temperature-induced volume contraction in GeTe},
  author={Chatterji, Tapan and Kumar, CMN and Wdowik, Urszula D},
  journal={Phys. Rev. B},
  volume={91},
  number={5},
  pages={054110},
  year={2015},
  publisher={APS}
}

@article{yashina2001xps,
  title={XPS study of fresh and oxidized GeTe and (Ge, Sn) Te surface},
  author={Yashina, LV and Kobeleva, SP and Shatalova, TB and Zlomanov, VP and Shtanov, VI},
  journal={Solid State Ion.},
  volume={141},
  pages={513--522},
  year={2001},
  publisher={Elsevier}
}

@article{sahu2024emergence,
  title={Emergence of giant orbital Hall and tunable spin Hall effects in centrosymmetric transition metal dichalcogenides},
  author={Sahu, Pratik and Bidika, Jatin Kumar and Biswal, Bubunu and Satpathy, S and Nanda, BRK},
  journal={Phys. Rev. B},
  volume={110},
  number={5},
  pages={054403},
  year={2024},
  publisher={APS}
}

@article{lee2021efficient,
  title={Efficient conversion of orbital Hall current to spin current for spin-orbit torque switching},
  author={Lee, Soogil and Kang, Min-Gu and Go, Dongwook and Kim, Dohyoung and Kang, Jun-Ho and Lee, Taekhyeon and Lee, Geun-Hee and Kang, Jaimin and Lee, Nyun Jong and Mokrousov, Yuriy and others},
  journal={Commun. Phys.},
  volume={4},
  number={1},
  pages={234},
  year={2021},
  publisher={Nature Publishing Group UK London}
}

@article{lyalin2023magneto,
  title={Magneto-optical detection of the orbital Hall effect in chromium},
  author={Lyalin, Igor and Alikhah, Sanaz and Berritta, Marco and Oppeneer, Peter M and Kawakami, Roland K},
  journal={Phys. Rev. Lett.},
  volume={131},
  number={15},
  pages={156702},
  year={2023},
  publisher={APS}
}

@article{yu2014switching,
  title="{Switching of perpendicular magnetization by spin--orbit torques in the absence of external magnetic fields}",
  author={Yu, Guoqiang and Upadhyaya, Pramey and Fan, Yabin and Alzate, Juan G and Jiang, Wanjun and Wong, Kin L and Takei, So and Bender, Scott A and Chang, Li-Te and Jiang, Ying and others},
  journal={Nat. Nanotechnol.},
  volume={9},
  number={7},
  pages={548--554},
  year={2014},
  publisher={Nature Publishing Group}
}

@article{liu2011spin,
  title="{Spin-torque ferromagnetic resonance induced by the spin Hall effect}",
  author={Liu, Luqiao and Moriyama, Takahiro and Ralph, DC and Buhrman, RA},
  journal={Phys. Rev. Lett.},
  volume={106},
  number={3},
  pages={036601},
  year={2011},
  publisher={APS}
}

@article{manchon2019current,
  title="{Current-induced spin-orbit torques in ferromagnetic and antiferromagnetic systems}",
  author={Manchon, Aurelien and {\v{Z}}elezn{\`y}, Jakub and Miron, Ioan M and Jungwirth, Tom{\'a}{\v{s}} and Sinova, Jairo and Thiaville, Andr{\'e} and Garello, Kevin and Gambardella, Pietro},
  journal={Rev. Mod. Phys.},
  volume={91},
  number={3},
  pages={035004},
  year={2019},
  publisher={APS}
}

@article{nan2020controlling,
  title={Controlling spin current polarization through non-collinear antiferromagnetism},
  author={Nan, Tianxiang and Quintela, Camilo X and Irwin, Julian and Gurung, Gautam and Shao, Ding-Fu and Gibbons, J and Campbell, N and Song, Kuyngjun and Choi, S-Y and Guo, Lu and others},
  journal={Nat. Commun.},
  volume={11},
  number={1},
  pages={4671},
  year={2020},
  publisher={Nature Publishing Group UK London}
}

@article{kumar2021large,
  title={Large damping-like spin--orbit torque and improved device performance utilizing mixed-phase Ta},
  author={Kumar, Akash and Sharma, Raghav and Ali Khan, Kacho Imtiyaz and Murapaka, Chandrasekhar and Lim, Gerard Joseph and Lew, Wen Siang and Chaudhary, Sujeet and Muduli, Pranaba Kishor},
  journal={ACS Appl. Electron. Mater.},
  volume={3},
  number={7},
  pages={3139--3146},
  year={2021},
  publisher={ACS Publications}
}

@article{bangar2022large,
author = {Bangar, Himanshu and Kumar, Akash and Chowdhury, Niru and Mudgal, Richa and Gupta, Pankhuri and Yadav, Ram Singh and Das, Samaresh and Muduli, Pranaba Kishor},
title = {Large Spin-To-Charge Conversion at the Two-Dimensional Interface of Transition-Metal Dichalcogenides and Permalloy},
journal = {ACS Appl. Mater. Interfaces},
volume = {14},
number = {36},
pages = {41598-41604},
year = {2022},

}

@article{shao2016strong,
  title="{Strong Rashba-Edelstein effect-induced spin--orbit torques in monolayer transition metal dichalcogenide/ferromagnet bilayers}",
  author={Shao, Qiming and Yu, Guoqiang and Lan, Yann-Wen and Shi, Yumeng and Li, Ming-Yang and Zheng, Cheng and Zhu, Xiaodan and Li, Lain-Jong and Amiri, Pedram Khalili and Wang, Kang L},
  journal={Nano Lett.},
  volume={16},
  number={12},
  pages={7514--7520},
  year={2016},
  publisher={ACS Publications}
}

@article{sinova2015spin,
  title="{Spin Hall effects}",
  author={Sinova, Jairo and Valenzuela, Sergio O and Wunderlich, J{\"o}rg and Back, CH and Jungwirth, T},
  journal={Rev. Mod. Phys.},
  volume={87},
  number={4},
  pages={1213},
  year={2015},
  publisher={APS}
}

@article{pratik,
  title = {Effect of the inversion symmetry breaking on the orbital Hall effect: A model study},
  author = {Sahu, Pratik and Bhowal, Sayantika and Satpathy, S.},
  journal = {Phys. Rev. B},
  volume = {103},
  issue = {8},
  pages = {085113},
  numpages = {11},
  year = {2021},
  month = {Feb},
  publisher = {American Physical Society},
  doi = {10.1103/PhysRevB.103.085113},
  url = {https://link.aps.org/doi/10.1103/PhysRevB.103.085113}
}

\end{document}